\documentclass[oribibl]{llncs}

\PassOptionsToPackage{hyphens}{url}
\usepackage[colorlinks = true,
  linkcolor = blue,
  citecolor = blue,
  urlcolor = blue]{hyperref}

\usepackage{url}
\usepackage{csquotes}
\usepackage{multirow}
\usepackage{amsmath}
\usepackage{amssymb}
\usepackage{graphicx}
\usepackage{tabularx}
\usepackage{booktabs}

\begin{document}

\title{Digital Sovereignty as a Quality \\ Attribute for Software Architectures}

\author{Blinded\orcidID{blinded} \\
  \email{blinded}}
\institute{Blinded}

\author{Jukka Ruohonen\inst{1}\orcidID{0000-0001-5147-3084} \and \\
  Justin Stark\inst{2}\orcidID{0009-0001-3211-5764}
  \and Scott Wilkie\inst{2} \and \\
  Mikkel Baun Kj{\ae}rgaard\inst{1}\orcidID{0000-0001-5124-744X}}

\institute{University of Southern Denmark, S\o{}nderborg \& Odense, Denmark \\
  \email\{\texttt{juk, mbkj\}@mmmi.sdu.dk} \and University of Technology Sydney \& Accenture \\ \email\{\texttt{justin.stark, scott.wilkie\}@uts.edu.au}}

\maketitle

\begin{abstract}
Digital sovereignty (DS) is an increasingly important concept and political
agenda throughout the world, including in the European Union (EU). However, the
concept is also regrettably vague. With this critical point in mind, the paper
presents an analysis of digital sovereignty as a quality attribute for software
architectures in the context of cloud computing and the EU's policy frameworks
for it. The analysis reveals that DS can be sharpened analytically by
conceptualizing it as a quality attribute. The analysis further demonstrates how
DS satisfies many of the classical properties of quality attributes for software
architectures, including their measurability and validation, the trade-offs they
involve, and the scenario-based methodology commonly used for analyzing them.
\end{abstract}


\section{Introduction}

Digital sovereignty is an important but rather vague strategic project and
narrative in many countries and territories throughout the world, including but
not limited to Europe. When the concept's theoretical underpinnings and
associated theorization are omitted for brevity, digital sovereignty has often
been conceptualized through dependencies~\cite{Burwell26, Ruohonen26ICEDEG}. The
world is interconnected. Most parts of it are connected with each
other. According to the mainstream DS \text{narrative---or} most common DS
narratives at least, these dependencies cause also vulnerabilities for countries
and their sovereignties, alliances, and unions.

The responses vary, too. These include anything and everything from blanket data
localization mandates, territorial legal guardrails against extraterritorial
legal mandates, and industrial policy all the way to national
security. Throughout the world, including Europe, the responses are also
evolving. In fact, recently, in June 2026, the European Commission (EC)
introduced a new Tech Sovereignty Package~\cite{EC26a}. This package establishes
the paper's practical relevance and timeliness. The paper has also novelty. Even
though a lot of research (see, e.g., \cite{Schneider16} as an exemplar) has been
done to assess business decision-making criteria for choosing a given CC
infrastructure for a deployment, DS has received limited---but not
non-existent~\cite{OperaMartins16}, attention as a \textit{criterion} in
academic research. Having said that, the recent policy responses have prompted a
renewed interest on DS in the CC context~\cite{Stark26}. The paper contributes
to this timely research on DS that goes beyond philosophy and theorization, and
beyond politics and political~science.

With these introductory points in mind, the paper presents an analysis of
digital sovereignty as a quality attribute (QA) for software
architectures~(SAs). To narrow the scope for feasibility reasons, the focus is
restricted to cloud computing~(CC) and the EU. With these framings and
restrictions of scope, the research question (RQ) examined is: How well the EU's
recent policy frameworks for CC align with a conceptualization of DS as a QA for
SAs? For answering to the questions, the paper builds methodologically upon a
combination of interpretative policy analysis~\cite{Browne19} and software
engineering literature on software architectures.

\section{Background}

\subsection{Quality Attributes}

Quality attributes are a classical way for designing and studying software
architectures. In software engineering more broadly, these resemble
non-functional requirements, although the demarcation is debated with respect to
functional requirements~\cite{Bass01}. Common QAs range from reliability,
security, and modifiability to performance, scalability, usability, and power
consumption---and everything in-between. When designing a software architecture,
QAs either exhibit or inhibit architectural design choices. They often also
exhibit or inhibit actual implementation work---or detailed design, as it is
sometimes known. If security is prioritized as a QA, for instance, a detailed
design may be constrained by coding standards, a preference of memory safe
programming languages, and security testing, among other things. At the same
time, architectural design choices may be constrained by attack surface
minimization and related considerations.

A good quality attribute should be \textit{verifiable} and, whenever possible,
\textit{measurable} or otherwise possible to empirically
evaluate~\cite{Bachmann05}. A classical way to achieve this evaluation
requirement is to consider different \textit{scenarios} for architectures and
deployments operating in given environments with different stimuli, sources of
these, responses to them, and response measures and metrics~\cite{Bachmann05,
  Bass01, Kjaergaard15}. If availability is of high importance, for instance, a
deviation from a ``24/7/363 baseline'' is rather easy to measure against stimuli
of hardware failures, denial of service attacks, or something alike. The point
is also good because some QAs, such as availability and performance, can only be
measured and evaluated at runtime, while some other QAs can be evaluated already
during development~\cite{IEEE25}. Another related point is that a
\textit{risk-based} approach aligns with and complements the scenario-based
viewpoint to quality attributes and software
architectures~\cite{Fairbanks10}. A~further important point is that many QAs
involve \textit{trade-offs} with other QAs~\cite{Barney12, Bass01,
  IEEE25}. Security may involve balancing with usability, privacy, and
performance, and performance in turn may involve a trade-off with power
consumption, and so forth. In~what follows, these points motivate the subsequent
consideration and conceptualization of digital sovereignty as a distinct QA.

\subsection{Conceptualizing DS as a QA for SAs}\label{subsec: conceptualizing}

The point in the introduction about DS having often been framed through
dependencies aligns well with SAs and software in general. To exclude some
outliers, practically all software contain some dependencies. There are also
different types of dependencies; some dependencies are needed to execute a
program, whereas others are required to develop it and some are only needed
during installation~\cite{Latendresse12}. With SAs, the level of abstraction is
higher also regarding dependencies. Among other things, it is possible to
consider \textit{semantic dependencies}, such as standards, protocols, file
formats, and meta-data schemas~\cite{Bass01}. Components of a SA may depend on
each other also through resources available for them or through \textit{temporal
  dependencies}. The two theoretical dependency types emphasized provide a
further framing for the paper and a further restriction to its scope: the focus
is on semantic dependencies instead of strictly technical software dependencies,
while temporal dependencies are proxied via the classical concept of vendor
\textit{lock-in}. Throughout the history of information technology, lock-in has
been a classical business strategy of many vendors. Whether it is other
businesses, public administrations, or consumers, the switching costs after a
successful lock-in are often too high to migrate away from a given
vendor~\cite{OECD25, Zhu12}.

Partially due to a lack of standardization and partially due to business
strategies of CC vendors, lock-in has been pronounced in the CC
domain~\cite{OperaMartins16}. For conceptualizing lock-in analytically via
temporal dependencies, a template is presented in Subsection~\ref{subsec:
  scenario} for \textit{ex~post} scenarios. That is to say, the focus is not on
\textit{ex~ante} business considerations about a given CC vendor but rather on
\textit{ex~post} considerations about migrating away from a given CC vendor
already~used. With respect to SA literature on QAs~\cite{Kjaergaard15}, the
template thus signifies \textit{adaptability} as a distinct QA in the context of
cloud computing and strategic decision-making.

Regarding deriving metrics for DS as a QA for SAs, it seems sensible to consider
a component-based architecture with components covering anything from source
code modules to data and algorithms. By following and adopting two concepts
presented in the SA literature~\cite{Bass01, Silva23}, it seems sensible to
consider the \textit{extent} of a SA's components being semantically dependent
on a CC vendor as well as the \textit{distance} between the dependencies. The
latter concept can be understood as a difficulty of resolving conflicts between
the dependencies.

To elaborate: if a SA has $n$ \textit{components}, $\mathcal{C}_1, \ldots,
\mathcal{C}_n$, and $m$ of them are deployed in a given \textit{cloud}
$\mathbb{C}_i$, a ratio $\alpha = m/n$ defines a simple metric for the semantic
dependence of the SA on the $i$:th cloud vendor's $\mathbb{C}_i$. Let $\beta$
further denote the \textit{strength of vendor lock-in}, $\beta = 0$ indicating
no lock-in power whatsoever and $\beta = 1$ extreme lock-in. When $\beta \to 1$,
the difficulties involved in a migration away from a given cloud
increase. Therefore, $\alpha \times \beta$ can analytically be seen to proxy an
overall difficulty of a migration strategy from a business perspective. With a
reference to the points later raised in Subsection~\ref{subsec: cc background},
let $\zeta \in [0, 1]$ further denote the \textit{extent of applicability of
  foreign jurisdictions} over EU law. When $\zeta = 0$, only EU law matters, and
the other way around. When intellectual property, trade secrets, and
confidentiality in general are further considered, a metric $\sigma = \alpha
\times \beta \times \zeta$ provides an analytical baseline for a risk
analysis regarding a~$\mathbb{C}_i$.

By following cyber security risk analysis literature~\cite{Allodi17}, it is
useful to further consider the probability of a given stimulus occurring and the
impact of the stimuli upon a business, including even business continuity as an
extreme case. If $\rho$ thus denotes the \textit{occurrence probability} and the
\textit{impact} too ranges from zero to one, $\mathcal{I} \in [0, 1]$, higher
values proxying severity, the overall equation to consider is $\phi = \sigma
\times \rho \times \mathcal{I}$. Although conditional probabilities could be
considered, it suffices to say, for brevity and simplicity, that the fundamental
business strategy would be to try to minimize $\phi$. While several options are
available, the obvious option would be to prefer entirely on-premise
deployments, such that $\alpha = 0$ and thus $\phi = 0$ too. But when $\alpha
\to 1$, the other parameters should be minimized. For instance, $\mathcal{I} \to
0$ could be attempted by legal contracting, meaning that $\zeta \to 0$.

\section{Cloud Computing as a Case}\label{sec: cc}

\subsection{Background}\label{subsec: cc background}

The Western cloud computing business is dominated by a few large multinational
companies headquartered in the United States (US). In fact, so dominant they are
that some have called them and their infrastructures as ``cloud
empires''~\cite{Lehdonvirta24}. Also the lock-in is strongly felt in
Europe~\cite{Rao26}. According to some observers, ``European governments and
enterprises are bound hand and foot to US cloud service providers'', and they
``rarely even manage to switch a service from one US supplier to another US
supplier'' \cite{Hubert25}. Given the notation that was introduced in
Subsection~\ref{subsec: conceptualizing}, it therefore seems reasonable to
analytically fix $\beta > 0.9$, say.

As has been argued previously~\cite{Ruohonen25DS}, for a customer of a CC
infrastructure, it does not necessarily matter that much where the
infrastructure is physically located. The same applies to extraterritorial legal
mandates, including, but hardly limited to, the CLOUD Act of the US. In other
words, an extraterritorial legal mandate of one country may apply to a CC
infrastructure physically located in another country. While the legal details
are difficult~\cite{Naef23}, extraterritoriality has long had also a political
and even a geopolitical aspect. For instance, it has been seen as the underlying
concept behind the notion of ``Brussels effect'', the promotion of European
values and exertion of soft power by the EU via
regulation~\cite{Gstrein21}. Recently, an analogous notion of ``Washington
effect'' was introduced for describing the exertion of harder power by the
US~\cite{Politico25}. Given these concepts and their political roots, a further
concept of ``sovereignty 2.0'' was recently initiated for separating the older,
territorial ``sovereignty 1.0'' from the newer ``sovereignty by control and
sovereignty by design''~\cite[p.~3]{Stark26}. Regarding sovereignty by design,
the parameter $\alpha = m/n$ would be a simple but illuminating
example. Another, more theoretical example would be sovereignty by design
through the notion of ``regulation by design''~\cite{Prifti24}. Regarding
sovereignty by control, a relevant question is about ``who can administer,
access, approve, evidence, and recover services''~\cite[p.~3]{Stark26}. These
concepts help at abstracting and theorizing also the new policy frameworks for
cloud computing in the European~Union.

\subsection{The Cloud Sovereignty Framework}\label{subsec: csf}

The cloud computing initiatives \textit{by} the EU, but not necessary
\textit{in} the EU as a whole, have included both infrastructure projects and
policy responses, including new legislative acts and informal guidelines. While
at the time of writing it seems that the pan-European GAI-X project for CC has
failed to deliver, the Euro\-HPC ecosystem has gained traction also in terms of
concrete supercomputing infrastructures~\cite{Ruohonen25DS}. Regarding policy
responses, is also seems that the European Cloud Certification Scheme (ECCS),
which was initiated in conjunction with the enactment of the Cybersecurity Act
in 2019, failed to deliver, partially due to political in-fighting in Europe and
partially due to external pressure~\cite{EUISS25, Burwell26, EC26c}. However, in
2025 the EC released an updated version for the so-called Cloud Sovereignty
Framework (CSF). It and the Tech Sovereignty Package proposed are particularly
well-suited for addressing the RQ specified in the introduction.

\begin{figure*}[th!b]
\centering
\includegraphics[width=\linewidth, height=8.5cm]{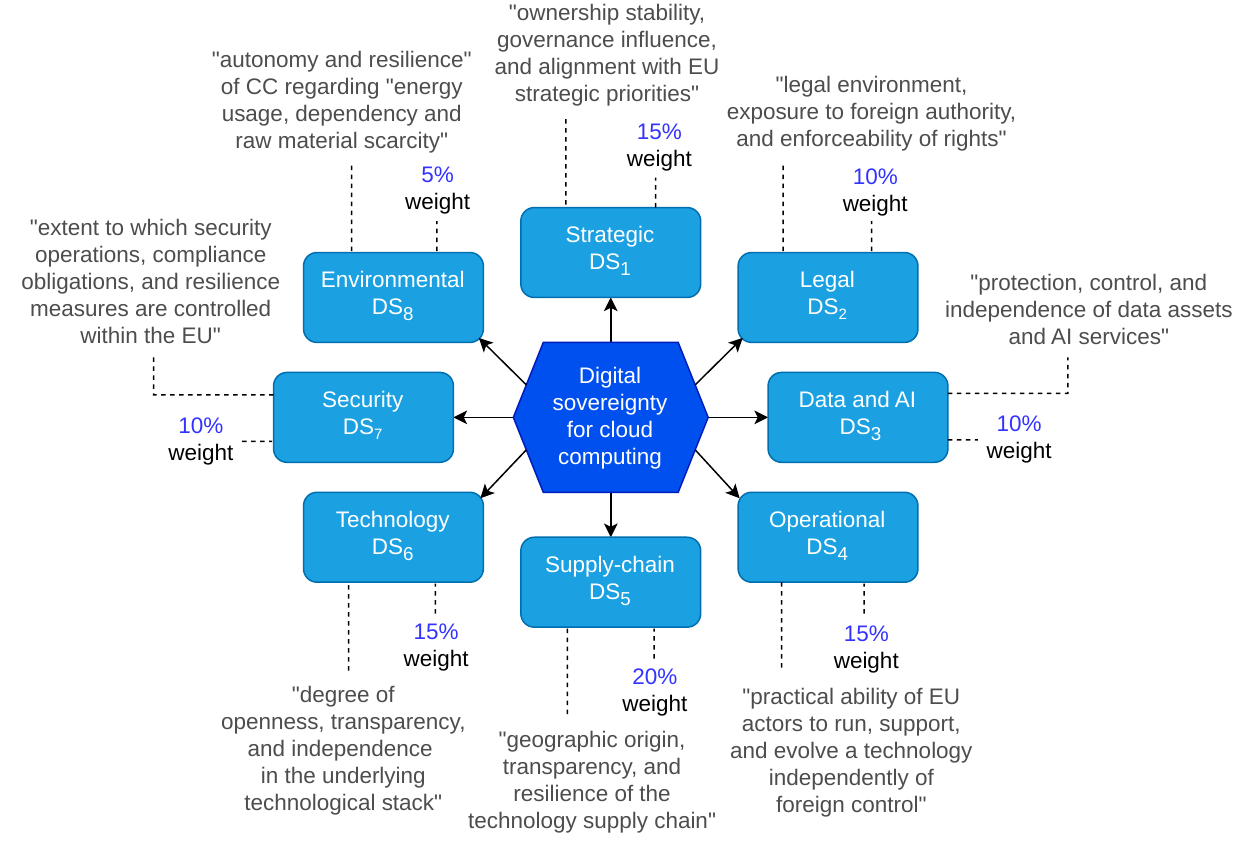}
\caption{The CSF's DS Breakdown (quotations and other references from
  \cite{EC25})}
\label{fig: csf}
\end{figure*}

The CSF specifies a five-fold effectiveness scale for DS, lower values
indicating less digital sovereignty and the maximum score denoting full digital
sovereignty, meaning that both technology and operations are ``under complete EU
control, subject only to EU law, with no critical non-EU
dependencies''~\cite[p.~3]{EC25}. This scale can be seen to provide an overall
key performance indicator. However, the gist of the CSF is the breakdown of DS
into eight distinct dimensions. Rather similar breakdowns have also been
recently done by think-tanks and their policy
analysts~\cite{Burwell26}. Regarding the CSF, Fig.~\ref{fig: csf} presents a
summary of its breakdown. Before continuing to disseminate it, should be
recalled the conceptualization derived uses a unit interval with lower values
indicating more digital sovereignty.

As can be seen, the CSF substantially extends the scope of DS particularly in
the context of \textit{public procurement} for which the CSF is primary
intended. Regarding the previous exposition and notation, it can be assumed that
the parameter $\zeta$ captures the $\textmd{DS}_2$ dimension. For not
overloading the notation further, let a scalar $\epsilon \in [0, 1]$ capture the
CSF's remaining dimensions, the set $S = \lbrace \textmd{DS}_1, \textmd{DS}_3,
\textmd{DS}_4, \textmd{DS}_5, \textmd{DS}_6, \textmd{DS}_7, \textmd{DS}_8
\rbrace$. When $\epsilon$ increases, the elements in $S$ either together or
individually contribute to decreasing digital sovereignty. For instance,
$\epsilon$ could attain a high value in case there are no raw materials required
to manufacture chips needed to build a new European $\mathbb{C}_i$. In this case
it would be primarily $\textmd{DS}_8$ and perhaps $\textmd{DS}_5$ secondarily
behind the high value~for~$\epsilon$.

What else can be analytically deduced from Fig.~\ref{fig: csf}?  Analogously to
the earlier fixing of $\beta$ to a high-value in Subsection~\ref{subsec: cc
  background}, it seems justified to predefine also $\epsilon$ to a relatively
high value. For instance, European actors have limited, if any, ability to
independently evolve a technology powering the overall technology stack of a
cloud $\mathbb{C}_i$. Having said that, the CSF is useful as an analytical
vehicle for comparing a $\mathbb{C}_i$ against another $\mathbb{C}_k$, $i \neq
k$. The same applies to the \textit{ex~post} migration strategies that were
motivated in Subsection~\ref{subsec: conceptualizing}. Regarding the
measurability preference for QAs, the CSF also comes with a formula for
comparisons~\cite[p.~6]{EC25}. With some alterations, the formula takes the form
of:
\begin{equation}
  s_i = \sum^{\vert S_i \vert}_{j \in S_i} \frac{f(\textmd{DS}^i_j) \times w^i_j}{\max(s_i, s_k) \times 100} , \quad \textmd{DS}^i_j \in S_i~\textmd{for}~\mathbb{C}_i,
\end{equation}
where $s_i \in [0, 1]$ is a score for a $\mathbb{C}_i$ and $s_k \in [0, 1]$ a
score for a $\mathbb{C}_k$, lower values indicating more sovereignty, $f(\cdot)
= y$ is a scoring function used by a human evaluator to rank a given DS
dimension, $y \in [0, 1]$, higher values indicating less sovereignty, and
$w^i_j$ is a weight from Fig.~\ref{fig: csf} for the $j$:th DS dimension of the
$i$:th cloud computing vendor. Thus: if $s_i < s_k$, $\mathbb{C}_i$ would be
preferable over $\mathbb{C}_k$ from a digital sovereignty perspective. That
said, the other parameters elaborated earlier should be considered too; a
similar function might be used also for~them. Finally, a scalar $\Phi = \phi
\times s$ could be used for summarizing the final yet still incomplete risk
equation; as previously, the subscripts are omitted for brevity. A~final point
is that a function would be needed to transform the values into the five-fold
effectiveness scale of the CSF. In practice, it might be easier to operate with
parameters defined in ordinal scale to begin with; the direct alignment with
$\rho$ justifies the preference of the unit interval for the present analytical
work.

\subsection{The Cloud and AI Development Act Proposal}\label{subsec: caida}

The EU's recent Tech Sovereignty Package contains also a legislative proposal
for a new Cloud and AI Development Act (CADA).\footnote{~If not otherwise
explicitly indicated with a different reference, the discussion in this
Subsection~\ref{subsec: caida} is from \cite[pp.~51--62]{EC26b}, including with
respect to direct quotations.} By excluding the artificial intelligence (AI)
part, the proposal's CC part continues the CSF from a legal perspective. It too
specifies assurance levels for CC, including with respect to their
self-assessments, independent third-party audits, transparency obligations, and
mandates for European public authorities. However, the proposal's cloud-specific
gist is on Articles~29 and 30 that specify requirements for risk analysis and
public procurement, respectively. According to the former article, both EU
institutions and the member states are required to carry out risk assessments
for public sector activities involving a use of CC services. According to
Article~29(2), they should pay particular attention to the following three
aspects at minimum:
\begin{enumerate}
\itemsep 5pt
\item{The \textit{data and its characteristics} stored to a cloud
  $\mathbb{C}_i$, including ``sensitivity, criticality, and magnitude of the
  non-personal data processed''. With respect to personal data, magnitude and
  risks to the rights and freedoms of people should be assessed, as is the case
  in general with data protection law.}
\item{The ``risk and consequent impact'' of ``\textit{unlawful access}'' under
  EU law.}
\item{The ``risk and consequent impact on public order'' for a ``\textit{service
    disruption}''.}
\end{enumerate}

Of these aspects, the first falls primarily to the domain of $\textmd{DS}_3$ in
the CSF and the second primarily to the domain of $\textmd{DS}_2$. The third
aspect is briefly discussed further in the forthcoming Subsection~\ref{subsec:
  scenario}. When compared to the ``all hazards'' approach present in a notable
cyber security directive~\cite{Ruohonen24ISJGP}, the CADA proposal's risk
assessment scope is perhaps surprisingly narrow with respect to CC. However,
there are two other particularly noteworthy things in Article~29.

First, Article~29(6) of the proposal specifies that the risk assessment results
are used for determining whether a public sector entity should switch from a
$\mathbb{C}_i$ to a $\mathbb{C}_k$ due to undue risk with the $i$:th cloud. If
that is determined to be the case, they should do the migration within twelve
months. Thus, the CADA proposal reiterates rather closely the earlier points
about migration. Given what was said about $\beta$ and $\epsilon$, a migration
may be easier said than done, however.

Second, Article~29(9) specifies that the risk assessments should also evaluate
``whether a multi-vendor or multi-cloud strategy is appropriate'' when procuring
new CC services. Even though cloud vendors actively try to disincentivize such
``multi-homing'' strategies~\cite{OECD25}, the evaluation seems particularly
relevant from a SA perspective. Given the earlier points about the scalar
$\alpha$, components from $\mathcal{C}_1$ to $\mathcal{C}_n$, two clouds
$\mathbb{C}_i$ and $\mathbb{C}_k$, and the impact $\mathcal{I}$ from a realized
risk, it may make sense to deploy some of the components in $\mathbb{C}_i$ and
some others in $\mathbb{C}_k$. Analogously to cyber security
reasoning~\cite{Garcia14, Hosseinzadeh15}, \textit{diversification} may thus be
a good idea for mitigating the risks identified and perhaps those unidentified
as well~\cite{Baldoni25}. From an architectural perspective, a given SA should
be well-decomposed for such a multi-cloud deployment to be possible. As has been
argued also with respect to policy measures~\cite{Bendinelli25, OECD25}, a sound
migration from a cloud $\mathbb{C}_i$ to a cloud $\mathbb{C}_k$ requires also
\textit{interoperability} between the two clouds. Given the $\textmd{DS}_1$
dimension, strategic decision-making might consider also other aspects, such as
using $\mathbb{C}_i$ from a CC vendor headquartered in one country and
$\mathbb{C}_k$ from another vendor headquartered in a different country. If
resources are not a concern, which might or should be the case with some highly
critical public sector services, \textit{redundancy} would be another
option~\cite{Baldoni25}; $\mathcal{C}_1, \ldots, \mathcal{C}_n$ could be
deployed in both $\mathbb{C}_i$~and~$\mathbb{C}_k$.

Finally, a brief remark is warranted about the CADA proposal's assurance levels
because they demonstrate how the DS concept is evolving. When going through the
Annex II of the proposal~\cite{EC26d} and using the terminology associated with
the ``sovereignty 2.0'' concept~\cite{Stark26}, the assurance levels one and two
address the ``data plane'' (such as location and processing), whereas the levels
three and four address the ``management plane'' (such as governance authority
and operational privileges). This terminological summary is worth to make
because it demonstrates how DS is evolving toward a sharper alignment with
technology.

\subsection{A Scenario Template}\label{subsec: scenario}

In what follows, a simple scenario template is presented for DS as a QA for SAs
in the CC context. It mimics the templates used in a well-known SA
textbook~\cite{Bass01}. Before continuing to briefly elaborate Fig.~\ref{fig:
  template}, it should be remarked that many of the aspects in the figure are
neither well-defined nor widely used---nor even yet well-understood at the time
of writing. Unlike with performance and scalability QAs for which concepts such
as latency and throughput are well-understood, even something like
time-to-migrate is difficult, if not impossible, to robustly assess in practical
terms---despite the limit of $12$ months in the CADA proposal.

\begin{figure*}[th!b]
\centering
\includegraphics[width=\linewidth, height=8.5cm]{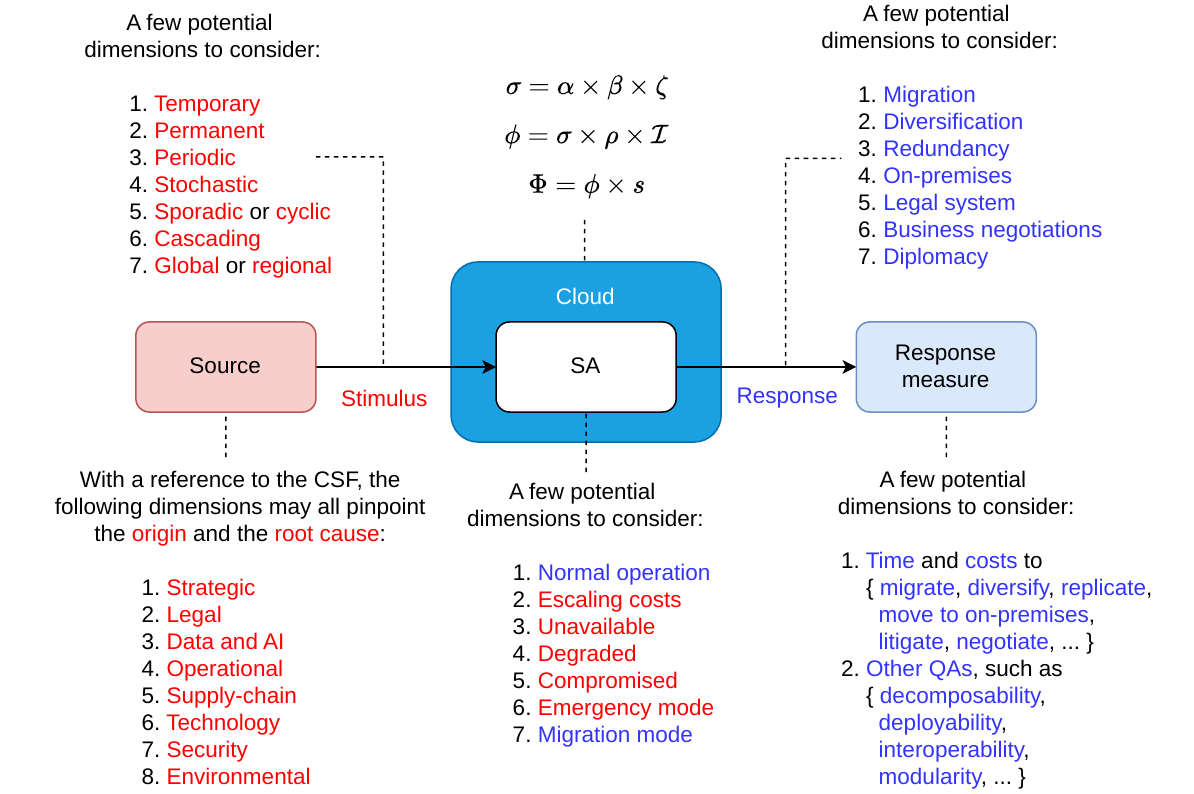}
\caption{A Simple Template for DS as a QA for SAs in the CC context}
\label{fig: template}
\end{figure*}

When the DS abbreviation is dropped from the earlier Fig.~\ref{fig: csf}, all
the eight dimensions can be seen either separately or jointly as a potential
source for a given stimulus. Even natural disasters sometimes disturb SAs and
other systems in some sectors~\cite[p.~14]{ENISA24}. Some of these may also be
periodic. Another example would come from the strategic $\textmd{DS}_1$
dimension and involve an acquisition of a $\mathbb{C}_i$ by a vendor. In such a
case it might be possible to negotiate the contracts and their guarantees with
the old vendor to apply also to the new vendor. If two cloud vendors merge, a
given $\Phi$ already evaluated should probably be re-assessed. In fact,
continuous evaluation would be beneficial, meaning that a subscript for time,
$t$, should probably be added, such that $\Phi_t, \Phi_{t+1}, \ldots$, and
likewise for the other parameters. Risks are not constant. In any case, these
examples demonstrate that the combinatorics in Fig.~\ref{fig: template} are
almost~infinite.

However, it has been geopolitics that have driven the European policy discussion
in recent years. These would originate from the strategic $\textmd{DS}_1$ of
another country. In this regard, both a possibility of degradation of services
and even a possibility of a ``kill switch'' have been speculated by
commentators~\cite{Rao26, Rudlph26}. Even though supposedly possible, $\rho$ is
likely minimal for someone pressing such a kill switch. Nevertheless, the
speculative example is worth noting because $\mathcal{I}$ would be
catastrophic. Of the potential dimensions listed in Fig.~\ref{fig: template}, a
cloud would be immediately unavailable for a SA, meaning that even migration
would not be possible. Thus, the earlier points about diversification and
redundancy extend also to backups. Of the response dimensions listed in the
figure, diplomacy would likely be the option for trying to resolve the pressing
of the switch. Even if normal operations would be restored eventually, meaning
that the temporary dimension would apply for the source, the costs would likely
be excessive even from a short unavailability, provided that the switch would be
pressed wholesale.

\section{Conclusion and Discussion}

The answer to the RQ specified in the introduction is positive: the EU's recent
policy frameworks for CC align rather well with a conceptualization of DS as a
QA for SAs operating in a cloud. Regarding the desirable properties of QAs, the
paper demonstrated that measurability is attainable. Even though somewhat
sketchy, the EU's CSF is a good example in this regard. Given the complexities
involved, including legal uncertainties involving multiple jurisdictions, robust
verification may still be challenging even with quantified measures. When
combining this uncertainty with a risk-based and a scenario-based approach to QA
evaluation, the common recommendation \cite{DiZio24} for using Delphi surveys or
focus groups seems sensible. A joint evaluation of the equations involving
$\sigma$, $\phi$, and~$\Phi$ could sharpen the risks and scenarios
involved. Further research is required about mapping these to the other
dimensions in the template shown in Fig.~\ref{fig: template}, including with
respect to responses and response measures. Even though not explicitly
discussed, it is a truism that there are trade-offs as well. For instance, the
performance, scalability, and availability QAs of hyperscaling clouds are hard
to beat technically. Yet, it is difficult to say whether these or business
reasons are a reason for the Europe's continuing lock-in to the few
US-based~cloud~vendors. Furthermore, it should be acknowledged that the stimulus
in Fig.~\ref{fig: template} could also denote a large-scale crisis---including
even a military conflict---in which case the migration response in the figure
might actually denote a move \textit{to} a global hyperscaling cloud. The war in
Ukraine is a good example in this regard~\cite{Zapotichny26}.

To put aside on-premise deployments, diversification, interoperability, and
redundancy were among the few noteworthy points raised about decreasing the real
or potential risks. Of these, interoperability has been promoted by legislative
acts in the EU, whereas competition policy could be a leverage against barriers
for multi-homing deployments in multiple clouds~\cite{OECD25}. Indeed,
``single-homing'' in a single cloud has been observed to be more relevant
business wise~\cite{Mangold25}. Diversification and redundancy cost. They also
increase complexity of a SA and its detailed design. Whether these and related
factors continue to outweigh the risks from decreased DS remain to be seen. At
the time of writing, it seems the pendulum is shifting particularly on the side
public sector services and public administrations, which is understandable due
to the criticality of some of their functions. The EU's Tech Sovereignty Package
and the CADA proposal are good examples in this regard. At the same time, they
also signify that DS is increasingly aligning also with \textit{industrial
  policy} in the European Union.

However, it remains to be seen whether mere risk assessments are enough to
decrease the lock-in of European public sector services and public
administrations on the global hyperscalers. In this regard, some have already
criticized the CADA proposal in that it only prohibits foreign clouds at the
highest assurance level and even at that level a strict prohibition remains
unclear~\cite{ARTICLE19, TechPolicyPress26}. This criticism is worth mentioning
because it exemplifies that also more aggressive viewpoints are expressed in
Europe. Another point is that the criticism may have been too hasty because no
one knows yet how the scoring will work in practice---or whether the CADA
proposal will even be accepted in its current form.\footnote{~When recalling the
earlier point in Subsection~\ref{subsec: csf} about the ECCS, it is worth
mentioning another line of criticism about a potential fragmentation across the
member states and their public authorities, which may lead to so-called forum
shopping~\cite{Hartmann26}.} For instance, non-EU cloud computing vendors may
score well on $\textmd{DS}_4$, $\textmd{DS}_6$, and $\textmd{DS}_7$ through
audit commitments and process controls, whereas $\textmd{DS}_2$ remains
structurally unsolved for higher assurance levels. This point reiterates the
earlier remark about a need for transformation functions and threshold values.

What about the debated and blurry concept of digital sovereignty itself? It
seems that the concept is both evolving and expanding, as has been the case also
with some other related concepts such as critical
infrastructure~\cite{Ruohonen24ISJGP}. Critical infrastructure is also important
to note in this context because it demonstrates the dependencies and their risks
well. A good example would be a cost-cutting plan in one European country to
move election data to a foreign cloud~\cite{Yle25}. Although the plan was
canceled due to a scandal that followed, it demonstrates the trade-offs at a
symbolic yet fundamental level: for saving little money, a country was planning
to lock-in even its democratic core to a foreign cloud. As is typically the case
with quality attributes~\cite{Bass01}, \textit{prioritization} according to the
criticality for a society would seem as a sensible recommendation for
decision-makers seeking to move critical infrastructure functions to a
\text{cloud---or away from~it.}

Conceptual clarity is a big problem in many domains of
study~\cite{Basheky25}. The same has traditionally applied also to digital
sovereignty as a concept. Even though an unambiguous definition for DS urged by
some~\cite{Rudlph26} still seems to be missing, the analytical framing has
sharpened somewhat in recent years. The CSF's eight different dimensions for
digital sovereignty are a good example in this regard. At the same time, it
seems that competing concepts are emerging. To exclude the strategic autonomy
concept, a further ``sovereignty 2.0 concept'' of \textit{digital autonomy} has
emerged. It has been characterized as the ``ability to operate independently,
regardless of whether disruptions originate at the geopolitical or vendor
level''~\cite{RedHat25}.\footnote{~As has been noted~\cite[p.~3]{EPC20}, the EC
has used digital sovereignty, technological sovereignty, digital autonomy, and
strategic autonomy interchangeably in a series of publications, white papers,
and press releases. Therefore, it should be emphasized that the quote presented
does not necessarily match the EC's intended meaning.} The keywords of continued
independent operations and disruptions further bring the resilience buzzword
onto the conceptual table. The conceptualization presented can also be
acknowledged as a limitation: even though it was summarized into the single
parameter $\Phi$, it could be argued that the dimensions of digital sovereignty
for CC in Fig.~\ref{fig: csf} each contain their own QAs, meaning that the noun
attribute in the paper's title should perhaps be in a plural form.

Finally, the paper also exemplifies that DS continues to be a core part of the
``tussle'' that was predicted by some authors already nearly $25$ years
ago~\cite{Clark02}. But is there anything new?  While DS has always been also
about politics, the intensified geopolitics among Western countries are new in
the digital era.

\bibliographystyle{splncs03}

\end{document}